\documentclass[12pt]{article}

\usepackage[letterpaper, left=1truein, right=1truein, top = 1truein, bottom = 1truein]{geometry}

\usepackage{amsgen,amsmath,amstext,amsbsy,amsopn,amssymb,amsthm,setspace}
\usepackage[round]{natbib}
\usepackage[usenames,dvipsnames]{color}
\usepackage{appendix}
\usepackage{multirow}

\usepackage{listings}

\usepackage[boxed, vlined, lined, commentsnumbered]{algorithm2e}

\usepackage{hyperref}

\usepackage{ifpdf}
\ifpdf
\usepackage[pdftex]{graphicx}
\else
\usepackage{graphicx}
\fi

\newcommand{\be}{\begin{eqnarray}}
\newcommand{\ee}{\end{eqnarray}}
\newcommand{\ben}{\begin{eqnarray*}}
\newcommand{\een}{\end{eqnarray*}}
\newcommand{\bee}{\begin{enumerate}}
\newcommand{\eee}{\end{enumerate}}
\newcommand{\bei}{\begin{itemize}}
\newcommand{\eei}{\end{itemize}}
\renewcommand{\u}{\mathbf{u}}
\renewcommand{\v}{\mathbf{v}}

\newcommand{\Ex}{\mathrm{E}}
\newcommand{\median}{\mathrm{median}}
\newcommand{\corr}{\mathrm{corr}}
\newcommand{\sign}{\mathrm{sign}}
\newcommand{\supp}{\mathrm{supp}}
\newcommand{\hU}{{\hat U}}
\newcommand{\hV}{{\hat V}}
\newcommand{\hD}{{\hat D}}
\newcommand{\hu}{{\hat\u}}
\newcommand{\hv}{{\hat\v}}
\newcommand{\hd}{{\hat d}}
\newcommand{\R}{\mathbb{R}}

\ifpdf
\DeclareGraphicsExtensions{.pdf, .jpg, .tif}
\else
\DeclareGraphicsExtensions{.eps, .jpg}
\fi

\title{A Sparse SVD Method for High-dimensional Data}
\author{Dan Yang, Zongming Ma and Andreas Buja
\thanks{Dan Yang is Ph.D. Candidate, Department of Statistics, The Wharton School, University of Pennsylvania, Philadelphia, PA 19104 (Email: danyang@wharton.upenn.edu). Zongming Ma is Assistant Professor, Department of Statistics, The Wharton School, University of Pennsylvania, Philadelphia, PA 19104 (Email: zongming@wharton.upenn.edu). Andreas Buja is The Liem Sioe Liong/First Pacific Company Professor of Statistics, Department of Statistics, The Wharton School, University of Pennsylvania, Philadelphia, PA 19104 (Email: buja.at.wharton@gmail.com)}}

\date{}

\begin{document}
\maketitle

\begin{abstract}
We present a new computational approach to approximating a large,
noisy data table by a low-rank matrix with sparse singular vectors.
The approximation is obtained from thresholded subspace iterations
that produce the singular vectors simultaneously, rather than
successively as in competing proposals.  We introduce novel ways to
estimate thresholding parameters which obviate the need for
computationally expensive cross-validation.  We also introduce a way
to sparsely initialize the algorithm for computational savings that
allow our algorithm to outperform the vanilla SVD on the full data
table when the signal is sparse.  A comparison with two existing
sparse SVD methods suggests that our algorithm is computationally
always faster and statistically always at least comparable to the
better of the two competing algorithms.
\end{abstract}



\section{Introduction}
\label{sec:intro}

Singular value decompositions (SVD) and principle component analyses
(PCA) are the foundations for many applications of multivariate
analysis.  They can be used for dimension reduction, data
visualization, data compression and information extraction by
extracting the first few singular vectors or eigenvectors; see, for
example, \citet{Alter+01}, \citet{Prasantha+07}, \citet{Huang09},
\citet{Thomasian+98}.  In recent years, the demands on multivariate
methods have escalated as the dimensionality of data sets has grown
rapidly in such fields as genomics, imaging, financial markets.  A
critical issue that has arisen in large datasets is that in very high
dimensional settings classical SVD and PCA can have poor statistical
properties (\citealt{ShabalinNoble10}, \citealt{Nadler09},
\citealt{Paul07}, and \citealt{JohnstoneLu09}).  The reason is that in
such situations the noise can overwhelm the signal to such an extent
that traditional estimates of SVD and PCA loadings are not even near
the ballpark of the underlying truth and can therefore be entirely
misleading.  Compounding the problems in large datasets are the
difficulties of computing numerically precise SVD or PCA solutions at
affordable cost.  Obtaining statistically viable estimates of
eigenvectors and eigenspaces for PCA on high-dimensional data has been
the focus of a considerable literature; a representative but
incomplete list of references is \citet{Lu02}, \citet{Zou+06},
\citet{Paul07}, \citet{PaulJohnstone07}, \citet{ShenHuang08},
\citet{JohnstoneLu09}, \citet{Shen+11}, \citet{Ma11}.  On the other
hand, overcoming similar problems for the classical SVD has been the
subject of far less work, pertinent articles being \citet{Witten+09},
\citet{Lee+10}, \citet{Huang09} and \citet{Allen+11}.

In the high dimensional setting, statistical estimation is not
possible without the assumption of strong structure in the data.  This
is the case for vector data under Gaussian sequence models
\citep{Johnstone11}, but even more so for matrix data which require
assumptions such as low rank in addition to sparsity or smoothness.
Of the latter two, sparsity has slightly greater generality because
certain types of smoothness can be reduced to sparsity through
suitable basis changes \citep{Johnstone11}.  By imposing sparseness on
singular vectors, one may be able to ``sharpen'' the structure in data
and thereby expose ``checkerboard'' patterns that convey biclustering
structure, that is, joint clustering in the row- and column-domains of
the data (\citealt{Lee+10} and \citealt{Sill+11}).  Going one step
further, \citet{WittenTibshirani10} used sparsity to develop a novel
form of hierarchical clustering.

So far we implied rather than explained that SVD and PCA approaches
are not identical.  Their commonality is that both apply to data that
have the form of a data matrix $X = (x_{ij})$ of size $n \times p$.
The main distinction is that the PCA model assumes the rows of $X$ to
be i.i.d.~samples from a $p$-dimensional multivariate distribution,
whereas the SVD model assumes the rows $i=1,2,...,n$ to correspond to
a ``fixed effects'' domain such as space, time, genes, age groups,
cohorts, political entities, industry sectors,~...~.  This domain is
expected to have near-neighbor or grouping structure that will be
reflected in the observations $x_{ij}$ in terms of smoothness or
clustering as a function of the row domain.  In practice, the
applicability of either approach is often a point of debate (e.g.,
should a set of firms be treated as a random sample of a larger domain
or do they constitute an enumeration of the domain of interest?), but
in terms of practical results the analyses are often interchangeable
because the points of difference between the SVD and PCA models are
immaterial in the exploratory use of these techniques.  The main
difference between the models is that the SVD approach analyzes the
matrix entries as structured {\em low-rank means plus error}, whereas
the PCA approach analyzes the covariation between the column
variables.


In modern developments of PCA, interest is focused on ``functional''
data analysis situations or on the analog of the ``sequence model''
\citep{Johnstone11} where the columns also correspond to a structured
domain such as space, time, genes,~...~.  It is only with this focus
that notions of smoothness and sparseness in the column or row domain
are meaningful.  A consequence of this focus is the assumption that
all entries in the data matrix have the same measurement scale and
unit, unlike classical PCA where the columns can correspond to
arbitrary quantitative variables with any mix of units.  With
identical measurement scales throughout the data matrix, it is
meaningful to entertain decompositions of the data into signal and
fully exchangeable noise:
\begin{equation} \label{eq:model}
	X = \Xi + Z\,,
\end{equation}
where $\Xi = (\xi_{ij})$ is an $n\times p$ matrix representing the
signal and $Z = (z_{ij})$ is an $n \times p$ random matrix
representing the noise and consisting of i.i.d.~errors as its
components.  In both PCA and SVD approaches, the signal is assumed to
have a multiplicative low-rank structure: $\Xi = U D V' = \sum_{l=1}^r
d_{l}\u_{l}\v_{l}'$, where for identifiability it is assumed that rank
$r < \min(n,p)$, usually even ``$\ll$'' such as $r=1,~2$ or $3$.  The
difference between SVD and PCA is, using ANOVA language, that in the
SVD approach both $U$ and $V$ represent fixed effects that can both be
regularized with smoothness or sparsity assumptions, whereas in
functional PCA $U$ is a random effect.  As indicated above, such
regularization is necessary for large $n$ and $p$ because for
realistic signal-to-noise ratios recovery of the true $U$ and $V$ may
not be possible.  ---~Operationally, estimation under sparsity is
achieved through thresholding.  In general, if both matrix dimensions
are thresholded, one obtains sparse singular vectors of $X$; if only
the second matrix dimension is thresholded, one obtains sparse
eigenvectors of $X'X$, which amounts to sparse PCA.

A few recent papers propose sparsity approaches to the high
dimensional SVD problem: \citet{Witten+09} introduced a matrix
decomposition which constrains the $l_1$ norm of the singular vectors
to impose sparsity on the solutions.  \citet{Lee+10} used penalized LS
for rank-one matrix approximations with $l_1$ norms of the singular
vectors as additive penalties.  Both methods use iterative procedures
to solve different optimization problems.  [We will give more details
  about these two methods in Section \ref{sec:simulation}.]
\citet{Allen+11} is a Lagrangian version of \citet{Witten+09} where
the errors are permitted to have a known type of dependence and/or
heteroscedasticity.  These articles focus on estimating the first
rank-one term given by $\hd_1, \hu_1, \hv_1$ by either constraining
the $l_1$ norm of $\hu_1$ and $\hv_1$ or adding it as a penalty.  To
estimate $\hd_2, \hu_2, \hv_2$ for a second rank-one term, they
subtract the first term $\hd_1\hu_1\hv_1'$ from the data matrix $X$
and repeat the procedure on the residual matrix.  There exists further
related work on sparse matrix factorization, for example, by
\citet{Zheng+07}, \citet{Mairal+10} and \cite{Bach+08}, but these do
not have the form of a SVD.  In our simulations and data examples we
use the proposals by \citet{Witten+09} and \citet{Lee+10} for
comparison.



Our approach is to estimate the subspaces spanned by the leading
singular vectors simultaneously.  As a result, our method yields
sparse singular vectors that are orthogonal, unlike the proposals by
\citet{Witten+09} and \citet{Lee+10}.  In terms of statistical
performance, simulations show that our method is competitive with the
better performing of the two proposals, which is generally
\citet{Lee+10}.  In terms of computational speed, our method is faster
by at least a factor of two compared to the more efficient of the two
proposals, which is generally \citet{Witten+09}.  Thus we show that
the current state of the art in sparse SVDs is ``inadmissible'' if
measured by the two metrics `statistical performance' and
`computational speed': our method closely matches the better
statistical performance and provides it at a fraction of the better
computational performance.  In fact, by making use of sparsity at the
initialization stage, our method also beats the conventional SVD in
terms of speed.

Lastly, our method is grounded in asymptotic theory that comprises
minimax results which we describe in a companion paper
(\citet{Yang+11}).  A signature of this theory is that it is not
concerned with optimization problems but with a class of iterative
algorithms that form the basis of the methodology proposed here.  As
do most asymptotic theories in this area, ours relies heavily on
Gaussianity of noise, which is the major aspect that needs
robustification when turning theory into methodology with a claim to
practical applicability.  Essential aspects of our proposal therefore
relate to lesser reliance on the Gaussian assumption.

The present article is organized as follows.  Section~\ref{sec:method}
describes our method for computing sparse SVDs.
Section~\ref{sec:simulation} shows simulation results to compare the
performance of our method with that of \citet{Witten+09} and
\citet{Lee+10}.  Section~\ref{sec:data} applies our and the competing
methods to real data examples.  Finally, Section~\ref{sec:discussion}
discusses the results and open problems.


\section{Methodology}
\label{sec:method}

In this section, we give a detailed description of the proposed sparse
SVD method.

To start, consider the noiseless case.  Our sparse SVD procedure is
motivated by the simultaneous orthogonal iteration algorithm
(\citealt{GolubLoan96}, Chapter 8), which is a straightforward
generalization of the power method for computing higher-dimensional
invariant subspaces of symmetric matrices.  For an arbitrary
rectangular matrix $\Xi$ of size $n\times p$ with SVD $\Xi = UDV'$,
one can find the subspaces spanned by the first $r$ ($1 \le r \le
\min(n,p)$) left and right singular vectors by iterating the pair of
mappings $V \mapsto U$ and $U \mapsto V$ with $\Xi$ and $\Xi'$ (its
transpose), respectively, each followed by orthnormalization, until
convergence.  More precisely, given a right starting frame $V^{(0)}$,
that is, a $p\times r$ matrix with $r$ orthonormal columns, the SVD
subspace iterations repeat the following four steps until convergence:
\begin{equation} \label{alg:vanilla}
\begin{array}{|ll|}
\hline
\mbox{(1) Right-to-Left Multiplication:}                  &U^{(k),mul}=\Xi V^{(k-1)} \\
\mbox{(2) Left Orthonormalization with QR Decomposition:} &U^{(k)}R_u^{(k)}=U^{(k),mul} \\
\mbox{(3) Left-to-Right Multiplication:}                  &V^{(k),mul}=\Xi' U^{(k)} \\
\mbox{(4) Right Orthonormalization with QR Decomposition:}&V^{(k)}R_v^{(k)}=V^{(k),mul} \\
\hline
\end{array}
\end{equation}
The superscript $^{(k)}$ indicates the $k$'th iteration, and $^{mul}$
the generally non-orthonormal intermediate result of multiplication.
For $r=1$, the QR decomposition step reduces to normalization.  If $\Xi$
is symmetric, the second pair of steps is the same as the first pair,
hence the original orthogonal iteration algorithm for symmetric
matrices is a special case of the above algorithm.


The problems our approach addresses are the following: For large noisy
matrices in which the significant structure is concentrated in a small
subset of the matrix $X$, the classical algorithm outlined above
produces estimates with large variance due to the accumulation of
noise from the majority of structureless cells \citep{ShabalinNoble10}.
In addition to the detriment for statistical estimation, involving large
numbers of structureless cells in the calculations adds unnecessary
computational cost to the algorithm.  Thus, shaving off cells with little
apparent structure has the promise of both statistical and computational
benefits.  This is indeed borne out in the following proposal for a
sparse SVD algorithm.



\subsection{The FIT-SSVD Algorithm: \\ ``Fast Iterative Thresholding for Sparse SVDs''}
\label{ssec:iter}

Unsurprisingly, the algorithm to be proposed here involves some form
of thresholding, be it soft or hard or something inbetween.  All
thresholding schemes reduce small coordinates in the singular vectors
to zero, and additionally such schemes may or may not shrink large
coordinates as well.  Any thresholding reduces variance at the cost of
some bias, but if the sparsity assumption is not too unrealistic, the
variance reduction will vastly outweigh the bias inflation.  The
obvious places for inserting thresholding steps are right after the
multiplication steps.  If thresholding reduces a majority of entries
to zero, the computational cost for the subsequent multiplication and
QR decomposition steps is much reduced as well.  The iterative
procedure we propose is schematically laid out in
Algorithm~\ref{algo:iter}.

\begin{algorithm}[tb]
\SetAlgoLined
\KwIn{}
1. Observed data matrix $X$. \\
2. Target rank~$r$. \\
3. Thresholding function $\eta$. \\
4. Initial orthonormal matrix $V^{(0)} \in \R^{p \times r}$. \\
5. Algorithm $f$ to calculate the threshold level $\boldsymbol{\gamma}
= f(X, U, V, \hat\sigma)$ given \\
(a)~the data matrix $X$, (b)~current
estimates of left and right singular vectors $U,\,V$, and (c)~an
estimate of the standard deviation of noise $\hat\sigma$. \\
(Algorithm \ref{algo:threshold-level} is one choice.)

\KwOut{Estimators $\hU = U^{(\infty)}$ and $\hV = V^{(\infty)}$.}
\nl Set $\hat\sigma=1.4826\,\mbox{MAD}\left(\mbox{as.vector}(X)\right)$. \\
\Repeat{Convergence}{
\nl Right-to-Left Multiplication: $U^{(k),mul} = X V^{(k-1)}$. \\
\nl Left Thresholding: $U^{(k),thr} = (u^{(k),thr}_{il})$, with
$u^{(k),thr}_{il}=\eta\left(u^{(k),mul}_{il},\gamma_{ul}\right)$,
\mbox{ where }$\boldsymbol{\gamma}_u=f(X, U^{(k-1)}, V^{(k-1)}, \hat\sigma)$. \\
\nl Left Orthonormalization with QR Decomposition: $U^{(k)}R_u^{(k)} = U^{(k),thr}$. \\
\nl Left-to-Right Multiplication: $V^{(k),mul} = X' U^{(k)}$. \\
\nl Right Thresholding: $V^{(k),thr} = (v^{(k),thr}_{jl})$, with
$v^{(k),thr}_{jl}=\eta\left(v^{(k),mul}_{jl},\gamma_{vl}\right)$,
\mbox{ where }$\boldsymbol{\gamma}_v = f(X', V^{(k-1)}, U^{(k)}, \hat\sigma)$. \\
\nl Right Orthonormalization with QR Decomposition: $V^{(k)}R_v^{(k)} = V^{(k),thr}$.
}
\caption{FIT-SSVD}
\label{algo:iter}
\end{algorithm}

In what follows we discuss the thresholding function and convergence
criterion of Algorithm~\ref{algo:iter}.  Subsequently, in
Sections~\ref{ssec:init}--\ref{ssec:thr}, we describe other important
aspects of the algorithm: the initialization of the orthonormal
matrix, the target rank, and the adaptive choice of threshold levels.

\paragraph{Thresholding function} At each thresholding step, we
perform entry-wise thresholding.  In our modification of the subspace
iterations~(\ref{alg:vanilla}) we allow any thresholding function
$\eta(x,\gamma)$ that satisfies $|\eta(x,\gamma)-x| \le \gamma$ and
$\eta(x,\gamma)1_{|x| \le \gamma}=0$, which includes soft-thresholding
with $\eta_{soft}(x,\gamma)=\sign(x)(|x|-\gamma)_+$, hard-thresholding
with $\eta_{hard}(x,\gamma)=x1_{|x|>\gamma}$, as well as the
thresholding function used in SCAD \citep{FanLi01}.  The parameter
$\gamma$ is called the threshold level.  In Algorithm~\ref{algo:iter},
we apply the same threshold level $\gamma_{ul}$ (or $\gamma_{vl}$) to
all the elements in the $l$th column of $U^{(k),mul}$ (or
$V^{(k),mul}$, resp.).  For more details on threshold levels, see
Section~\ref{ssec:thr}.

\paragraph{Convergence criterion}
We stop the iterations once subsequent updates of the orthonormal
matrices are very close to each other.  In particular, for any matrix
$H$ with orthonormal columns (that is, $H'H=I$), let $P_H = HH'$ be
the associated projection matrix.  We stop after the $k$th iteration
if $\max\{\|P_{U^{(k)}}-P_{U^{(k-1)}}\|_2^2,
\|P_{V^{(k)}}-P_{V^{(k-1)}}\|_2^2\} \leq \epsilon$, where $\epsilon$
is a pre-specified tolerance level, chosen to be $\epsilon = 10^{-8}$
for the rest of this article.  [$\|A\|_2$ denotes the spectral norm
  of~$A$.]


\subsection{Initialization algorithm for FIT-SSVD}
\label{ssec:init}

In Algorithm~\ref{algo:iter}, we need a starting frame $V^{(0)}$ such
that the subspace it spans has no dimension that is orthogonal to the
subspace spanned by the true $V$.  Most often used is the $V$ frame
provided by the ordinary SVD.  However, due to its denseness,
computational cost and inconsistency \citep{ShabalinNoble10}, it makes
an inferior starting frame.  Another popular choice is initialization
with a random frame, which, however, is often nearly orthogonal to the
true $V$ and thus requires many iterations to accumulate sufficient
power to converge.  We propose therefore Algorithm~\ref{algo:init}
which overcomes these difficulties.

\begin{algorithm}[tb]
\SetAlgoLined
\SetAlgoVlined
\KwIn{}
1. Observed data matrix~$X$. \\
2. Target rank~$r$. \\
3. Degree of ``Huberization'' $\beta$ (typically 0.95 or 0.99), \\
   \hspace{.2in}that defines a quantile of the absolute values of entries in~$X$. \\
4. Significance level of a selection test~$\alpha$.

\KwOut{Orthornormal matrices $\hU = U^{(0)}$ and $\hV = V^{(0)}$.}
\nl Subset selection: \label{algo:step:selection}\\
    Let $\delta$ be the $\beta$-quantile of the absolute values of all the entries in~$X$. \\
    Define $Y=(y_{ij})$ by $y_{ij}=\rho(x_{ij},\delta)$, where $\rho(x,\delta)$ is the Huber $\rho$ function: \\
    \hspace{.2in}$\rho(x,\delta)=x^2$ if $|x|\le \delta$ and $2\delta|x|-\delta^2$ otherwise. \\
    Select a subset $I=\{i_1,i_2,...\}$ of rows according to the next four steps:\\
    $-$ Let $t_{i}=\sum_{j=1}^{p} y_{ij}$ for $i = 1, \dots, n$. \\
    $-$ Let $\hat\mu=\mbox{median}(t_{1},...,t_{n})$ and $\hat{s}=1.4826\mbox{ MAD}(t_{1},...,t_{n})$. \\
    $-$ Calculate p-values: $p_{i}=1-\Phi(\frac{t_{i}-\hat\mu}{\hat{s}})$, where $\Phi$ is the CDF of $N(0,1)$. \\
    $-$ Perform the Holm method on the p-values at family-wise error rate $\alpha$, \\
    \hspace{.2in}and let $I$ be the indices of the p-values that result in rejection. \\
    Select a subset of columns $J$ similarly. \\
    Form the submatrix $X_{IJ}$ of size $|I| \times |J|$. \\
\nl Reduced SVD: Compute $r$ leading pairs of singular vectors of the submatrix $X_{IJ}$.\label{algo:step:svd} \\
    Denote them by $\u_1^I, \ldots, \u_r^I$ ($|I| \times 1$ each) and $\v_1^J, \ldots, \v_r^J$ ($|J| \times 1$ each). \\
\nl Zero-padding: Create $U^{(0)}=[\u_1^{(0)}, \ldots, \u_r^{(0)}]$ ($n \times r$) and
                         $V^{(0)}=[\v_1^{(0)}, \ldots, \v_r^{(0)}]$ ($p \times r$),\\
    such that $\u_{Il}^{(0)}=\u_{Il}^I\,,\; \u_{I^cl}^{(0)}=0\,, \; \v_{Jl}^{(0)}=\v_{Jl}^J\,,\; \v_{J^cl}^{(0)}=0$.
    \label{algo:step:zero-padding}
\caption{Initialization algorithm for FIT-SSVD.}
\label{algo:init}
\end{algorithm}

The algorithm is motivated by \citet{JohnstoneLu09} who obtained a
consistent estimate for principal components under a sparsity
assumption by initially reducing the dimensionality.  We adapt their
scheme to the two-way case, and we weaken its reliance on the
assumption of normal noise which in real data would result in too
great a sensitivity to even slightly heavier tails than normal.  To
this end we make use of some devices from robust estimation.  The
intent is to perform a row- and column-preselection
(Step~\ref{algo:step:selection}) before applying a classical SVD
(Step~\ref{algo:step:svd}) so as to concentrate on a much smaller
submatrix that contains much of the signal.  We discuss the row
selection process, column selection being analogous.

Signal strength in rows would conventionally be measured under
Gaussian assumptions with row sums of squares and tested with $\chi^2$
tests with $p$ degrees of freedom.  As mentioned this approach turns
out to be much too sensitive when applied to real data matrices due to
isolated large cells that may stem from heavier than normal tails.  We
therefore mute the influence of isolated large cells by Huberizing the
squares before forming row sums.  We then form approximate $z$-score
test statistics, one per row, drawing on the CLT since we assume $p$
(the number of entries in each row) to be large.  Location and scale
for the $z$-scores are estimated with the median and MAD (``median
absolute deviation'', instead of mean and standard deviation) of the
row sums, the assumption being that over half of the rows are
approximate ``null rows'' with little or no signal.  If the signal is
not sparse in terms of rows, this procedure will have low power, which
is desirable because it biases the initialization of the iterative
Algorithm~\ref{algo:iter} toward sparsity.  Using robust $z$-score
tests has two benefits over $\chi^2$ tests: they are robust to
isolated large values, and they avoid the sensitivity of $\chi^2$
tests caused by their rigid coupling of expectation and variance.
Finally, since $n$ tests are being performed, we protect against
over-detection due to multiple testing by applying Holm's
(1979)\nocite{Holm79} stepwise testing procedure at a specified
family-wise significance level~$\alpha$ (default: 5\%).  The end
result are a set of indices $I$ of ``significant rows''.  ---~The same
procedure is then applied to the columns, resulting in an index set
$J$ of ``significant columns''.

The submatrix $X_{IJ}$ is then submitted to an initial reduced SVD.
It is this initial reduction that allows the present algorithm to be
faster than a conventional SVD of the full matrix $X$ when the signal
is sparse.  The left and right singular vectors are of size $|I|$ and
$|J|$, respectively.  To serve as initializations for the iterative
Algorithm~\ref{algo:iter}, they are expanded and zero-padded to length
$n$ and $p$, respectively (Step~\ref{algo:step:zero-padding}).
---~This concludes the initialization Algorithm~\ref{algo:init}.


\subsection{Rank estimation}
\label{ssec:rank}

In Algorithm~\ref{algo:iter}, a required input is the presumed rank of
the signal underlying $X$.  In practice, we need to determine the rank
based on the data.  Proposals for rank estimation are the subject of a
literature with a long history, of which we only cite \citet{Wold78},
\citet{Gabriel02}, and \citet{Hoff07}.  The proposal we chose is the
bi-cross-validation (BCV) method by \citet{OwenPerry09}, but with a
necessary twist.

The original BCV method was proposed for low-rank matrices with dense
singular vectors.  Thus, we apply it to the submatrix $X_{IJ}$
obtained from the initialization Algorithm \ref{algo:init}, instead of
$X$ itself.  The submatrix should have much denser singular vectors
and, even more importantly, much higher signal to noise ratio compared
to the full matrix.  In simulations not reported here but similar to
those of Section \ref{sec:simulation}, BCV on $X_{IJ}$ yielded
consistent rank estimation when the signal was sufficiently strong for
detection in relation to sparsity and signal-to-noise ratio.



\subsection{Threshold levels}
\label{ssec:thr}

The tuning parameters $\gamma$ in the thresholding function
$\eta(x,\gamma)$ are called ``threshold levels''; they play a key role
in the procedure.  At each thresholding step in
Algorithm~\ref{algo:iter}, a (potentially different) threshold level
needs to be chosen for each column $l=1,...,r$ of $U^{(k)}$ and
$V^{(k)}$ to strike an acceptable bias-variance tradeoff.  In what
follows, we focus on $U^{(k)}$, while the case of $V^{(k)}$ can be
obtained by symmetry.

The goal is to process the iterating left and right frames in such a
way as to retain the coordinates with high signal and eliminate those
with low signal.  To be more specific, we focus on one column
$\u_l^{(k),mul} = X \v_l^{(k-1)}$.  Recall that $X$ is assumed to
admit an additive decomposition into a low-rank signal plus noise
according to model \eqref{eq:model}.  Then a theoretically sensible
(though not actionable) threshold level for $\u_l^{(k),mul}$ would be
$\gamma_{ul} = \Ex[ \|Z\v_l^{(k-1)}\|_{\infty}]$, where $Z$ is the
additive noise matrix, and $\|Z\v_l^{(k-1)}\|_{\infty}$ is the maximum
absolute value of the $n$ entries in the vector $Z\v_l^{(k-1)}$.  The
signal of any coordinate in $\u_l^{(k),mul}$ with value less than
$\gamma_{ul}$ could be regarded low since it is weaker than the
expected maximum noise level in the $l$'th rank given that there are
$n$ rows.

The threshold $\gamma_{ul}$ as written above is of course not
actionable because it involves knowledge of $Z$, but we can obtain
information by leveraging the (presumably large) part of $X$ that is
estimated to have no or little signal.  This can be done as follows:
Let $L_u$ be the index set of rows which have all zero entries in
$U^{(k-1)}$, and let $H_u$ be its complement; define $L_v$ and $H_v$
analogously.  We may think of $L_u$ and $L_v$ as the current estimates
of low signal rows and columns.  Consider next a reordering and
partitioning of the rows and columns of $X$ according to these index
sets:
\begin{eqnarray} \label{eq:partition}
X=
\begin{pmatrix}
	X_{H_uH_v}&X_{H_uL_v}\\
	X_{L_uH_v}&X_{L_uL_v}\\
\end{pmatrix}.
\end{eqnarray}
Since the entries in $\v_l^{(k-1)}$ corresponding to $L_v$ are zero,
only $X_{:H_v}$ (of size $n \times |H_v|$, containing the two left
blocks in (\ref{eq:partition})) is effectively used in the
right-to-left multiplication of the iterative
Algorithm~\ref{algo:iter}.  We can therefore simulate a ``near-null''
situation in this block by filling it with random samples from the
bottom right block which we may assume to have no or only low signal:
$X_{L_u L_v} \approx Z_{L_u L_v}$.  Denote the result of such
``bootstrap transfer'' from $X_{L_u L_v}$ to $X_{:H_v}$ by
$\tilde{Z}^*$ ($n \times |H_v|$).  Passing $\tilde{Z}^*$ through the
right-to-left multiplication with $\v_l^{(k-1)}$ we form $Z^* \v_{H_v
  l}^{(k-1)}$, which we interpret as an approximate draw from $Z
\v_l^{(k-1)}$.  We thus estimate $\|Z \v_l^{(k-1)}\|_{\infty}$ with
$\|Z^* \v_{H_v l}^{(k-1)}\|_{\infty}$, and $\Ex[\|Z
  \v_l^{(k-1)}\|_{\infty}]$ with a median of $\|Z^* \v_{H_v
  l}^{(k-1)}\|_{\infty}$ over multiple bootstraps of~$Z^*$.

In order for this  to be valid, the block $X_{L_u L_v}$ needs
to be sufficiently large in relation to $X_{:H_v}$.  This is the
general problem of the ``$m$ out of $n$'' bootstrap, which was
examined by \citet{Bickel+97}.  According to their results, this
bootstrap is generally consistent as long as $m=o(n)$.  Hence, when
the size $|L_u||L_v|$ of the matrix $X_{L_uL_v}$ is large, say, larger
than $n|H_v|\log(n|H_v|)$, we estimate
$\Ex[\|Z\v_1^{(k-1)}\|_{\infty}]$ by the median of $M$ bootstrap
replications for sufficiently large~$M$.  When the condition is
violated, $|H_v|$ tends to be large, the central limit theorem takes
effect, and each element of $Z\v_1^{(k-1)}$ would be close to a normal
random variable.  Thus, the expected value of the maximum is near the
asymptotic value $\sqrt{2\log{n}}$ times the standard deviation.
---~We have now fully defined the threshold $\gamma_{ul}$ to be used on
$\u_l^{(k),mul}$.  The thresholds for $l=1,...,r$ are then collected
in the threshold vector $\boldsymbol{\gamma}_u =
(\gamma_{u1},...,\gamma_{ur})'$.

A complete description of the scheme is given in
Algorithm~\ref{algo:threshold-level}.  Based on an extensive
simulation study, setting the number of bootstrap replications to $M =
100$ yields a good balance between the accuracy of the threshold level
estimates and computational cost.

\begin{algorithm}[tb]
\SetAlgoLined
\SetAlgoVlined
\KwIn{}
1. Observed data matrix $X\in \R^{n \times p}$\;
2. Previous estimators of singular vectors $U^{(k)} \in \R^{n \times r}$, $V^{(k)} \in \R^{p \times r}$\;
3. Pre-specified number $M$ of bootstraps\;
4. Estimate of the standard deviation of noise $\hat\sigma$.

\KwOut{Threshold level $\boldsymbol{\gamma}\in\mathbb{R}^r$.}
\nl Subset selection: $L_u=\{i: u^{(k)}_{i1} = ... = u^{(k)}_{ir} = 0\}$,
                      $L_v=\{j: v^{(k)}_{j1} = ... = v^{(k)}_{jr} = 0\}$, \\
   \hspace{1.2in}      $H_u=L_u^c, ~ H_v=L_v^c$\;
\nl \eIf{$|L_v||L_u|<n|H_v|\log(n|H_v|)$\label{algo:step:error-if-condition}}{
\nl \Return $\boldsymbol{\gamma} =
             \hat\sigma\sqrt{2\log(n)}\mathbf{1} \in \R^r$ \label{algo:step:error-normal}\;
}{
\nl \For{$i \leftarrow 1$ \KwTo $M$}{
\nl Sample $n|H_v|$ entries from $X_{L_uL_v}$ and reshape them into a matrix
    $\tilde{Z} \in \R^{n \times |H_v|}$\;
\nl $B=\tilde{Z}V_{H_v:}^{(k)} \in \R^{n\times r}$\;
\nl $C_{i:}=(\|B_{: 1}\|_\infty, \|B_{: 2}\|_\infty, \dots, \|B_{: r}\|_\infty)'$\;
}
\nl $\gamma_l = \median(C_{: l})$\;
\nl \Return $\boldsymbol{\gamma}=(\gamma_1,...,\gamma_r)'$.
}
\caption{The threshold level function $f(X,U,V,\hat{\sigma})$ for Algorithm~\ref{algo:iter}.
  As shown, the code produces thresholds for $U$.  A call to $f(X',V,U,\hat{\sigma})$
  produces thresholds for $V$.}
\label{algo:threshold-level}
\end{algorithm}



\subsection{Alternative methods for selecting threshold levels}
\label{ssec:thr-alt}

In methods for sparse data, one of the most critical issues is
selecting threshold levels wisely.  Choosing thresholds too small
kills off too few entries and retains too much variance, whereas
choosing them too large kills off too many entries and introduces too
much bias.  To navigate this bias-variance trade-off, we adopted in
Section~\ref{ssec:thr} an approach that can be described as a form of
testing: we established max-thresholds that are unlikely to be
exceeded by any $U$- or $V$-coordinates under the null assumption of
absent signal in the corresponding row or column of the data matrix.

To navigate bias-variance trade-offs, other commonly used approaches
include various forms of cross-validation, a version of which we
adopted for the different problem of rank selection in
Section~\ref{ssec:rank} (bi-cross-validation or BCV according to
\citet{OwenPerry09}).  Indeed, a version of cross-validation for
threshold selection is used by one of the two competing proposals with
which we compare ours: \citet{Witten+09} leave out random subsets of
the entries in the data matrix, measure the differences between the
fitted values and the original values for those entries, and choose
the threshold levels that minimize the differences.  Alternatively one
could use bi-cross-validation (BCV) for this purpose as well, by
leaving out sets of rows and columns and choosing the thresholds that
minimize the discrepancy between the hold-out and the predicted
values.  However, this would be computationally slow for simultaneous
minimization of two threshold parameters.  Moreover, the possible
values of the thresholds vary from zero to infinity, which makes it
difficult to choose grid points for the parameters.  In order to avoid
such issues, \citet{Lee+10} implement their algorithm by embedding the
optimization of the choice of the threshold level inside the
iterations that calculate $\u_l$ for fixed $\v_l$ and $\v_l$ for fixed
$\u_l$ (unlike our methods, theirs fits one rank at a time).  They
minimize a BIC criterion over a grid of order statistics of current
estimates.  This idea could be applied to our simultaneous
space-fitting approach, but the simulation results in
Section~\ref{sec:simulation} below show that the method of
\citet{Lee+10} is computationally very slow.

\section{Simulation results}
\label{sec:simulation}

In this section, we show the results of numerical experiments to
compare the performance of FIT-SSVD with two state-of-the-art sparse
SVD methods from the literature (as well as with the ordinary SVD).
In contrast to FIT-SSVD which acquires whole subspaces spanned by
sparse vectors simultaneously, both comparison methods are stepwise
procedures that acquire sparse rank-one approximations
$\hat{d}_l\hat\u_l\hat\v'_l$ successively; for example, the second
rank-one approximation $\hat{d}_2\hat\u_2\hat\v'_2$ is found by
applying the same method to the residual matrix $X-\hat{d}_1 \hat\u_1
\hat\v'_1$, and so on.  For both methods it is therefore only
necessary to describe how they obtain the first rank-one term.

\begin{itemize}

\item The first sparse SVD algorithm for comparison was proposed by
  \citet{Lee+10} [referred to from here on by their initials,
    ``LSHM''].  They obtain a first pair of sparse singular vectors by
  finding the solution to the following $l_1$ penalized SVD problem
  under an $l_2$ constraint:
  \begin{align*}
    \min_{\u,\v,s} \,\left( \|X-s\u\v'\|_F^2 + s\lambda_u\|\u\|_1 + s\lambda_v\|\v\|_1 \right).
    ~~~~~~{\rm subject~to}~~~~ \|\u\|_2 = \|\v\|_2 = 1.
  \end{align*}
  LSHM solve this problem by alternating between the following steps
  till convergence:
  \begin{eqnarray*}
    \begin{array}{ll}
      \mbox{(1)~~ }\tilde\u_l=\eta_{soft}(X\v_l^{old},\lambda_u)\,,
      & \u_l^{new}\leftarrow \frac{\tilde\u_l}{\|\tilde\u_l\|_2}\,,\\
      \mbox{(2)~~ }\tilde\v_l=\eta_{soft}(X'\u_l^{new},\lambda_v)\,,
      & \v_l^{new}\leftarrow \frac{\tilde\v_l}{\|\tilde\v_l\|_2}\,.\\
    \end{array}
  \end{eqnarray*}

\item The second sparse SVD algorithm for comparison with our proposal
  is the adaptation of the penalized matrix decomposition scheme by
  \citet{Witten+09} to the sparse SVD case [referred to as ``PMD-SVD''
    from here on].  They obtain the first pair of sparse singular
  vectors by imposing simultaneous $l_1$ and $l_2$ constraints on both
  vectors:
  \begin{eqnarray*}
    \min \|X-d\u\v'\|_F^2\,,\;\;\mbox{ subject to } \|\u\|_2=\|\v\|_2=1, \|\u\|_1\le s_u, \|\v\|_1\le s_v\,.
  \end{eqnarray*}
  The PMD-SVD algorithm iterates between the following two steps until convergence:
  \begin{eqnarray*}
    &&\mbox{(1) }\u=\frac{\eta_{soft}(X\v,\delta_u)}{\|\eta_{soft}(X\v,\delta_u)\|_2}\,,
    \mbox{ where } \delta_u \mbox{ is chosen by binary search such that } \|\u\|_1=s_u\,,\\
    &&\mbox{(2) }\v=\frac{\eta_{soft}(X'\u,\delta_v)}{\|\eta_{soft}(X'\u,\delta_v)\|_2}\,,
    \mbox{ where } \delta_v \mbox{ is chosen by binary search such that }
    \|\v\|_1=s_v\,.
  \end{eqnarray*}
\end{itemize}

To make fair comparisons, we use the implementations by their original
authors for both LSHM \citep{Lee+10code} and PMD-SVD
\citep{pma-package}.  The tuning parameters are always chosen
automatically by the default methods in their implementations.  For
FIT-SSVD, we always use $\eta=\eta_{hard}$ in Algorithm
\ref{algo:iter}, Huberization $\beta=0.95$ and Holm family-wise error
rate $\alpha=0.05$ in Algorithm \ref{algo:init}, and $M=100$
bootstraps in Algorithm \ref{algo:threshold-level}.  We did try
different values of $\alpha$, $\beta$ and $M$ in FIT-SSVD, and the
results are not sensitive to these choices.  Thus, in our experience
there is no need for cross-validated selection of these parameters.

In what follows, we report simulation results for situations in which
the true underlying matrix has rank one and two, respectively.
Throughout this section, the rank of the true underlying matrix is
assumed known.


\subsection{Rank-one results}
\label{ssec:simulation-r1}

In this part, we generate data matrices according to model
(\ref{eq:model}) with rank $r=1$, $n=1024$ and $p = 2048$, the
singular value $d_1$ ranging in $\{50, 100, 200\}$, and iid noise
$Z_{ij} \sim (\mu$=0,~$\sigma^2$=1).  At first glance $d_1=50$ may
appear like an outsized signal strength, but it actually is not: The
expected sum of squares of noise is $\Ex[\|Z\|^2_F] = np \approx $ 2
million, whereas the sum of squares of signal is a comparably
vanishing $\|d_1\u_1 \v_1'\|_F^2 = d_1^2 = 2500$, for a
signal-to-noise ratio $S/N = 0.0012$ (which makes the failure of the
ordinary SVD in these tasks less surprising).  Even $d_1=200$ amounts
to a $S/N = 0.012$ only.

As mentioned in the introduction, the FIT-SSVD method was motivated by
theoretical results that were based on Gaussian assumptions
\citep{Yang+11}; it is therefore a particular concern to check the
robustness of the method under noise with heavier tails than Gaussian.
To this end we report simulation results both for $N(0,1)$ and
${\textstyle \sqrt{3/5}}\,t_5$ noise, the latter also having unit
variance (the purpose of the factor ${\textstyle \sqrt{3/5}}$).

\begin{figure}[htp]
\centering \includegraphics[width=0.85\textwidth]{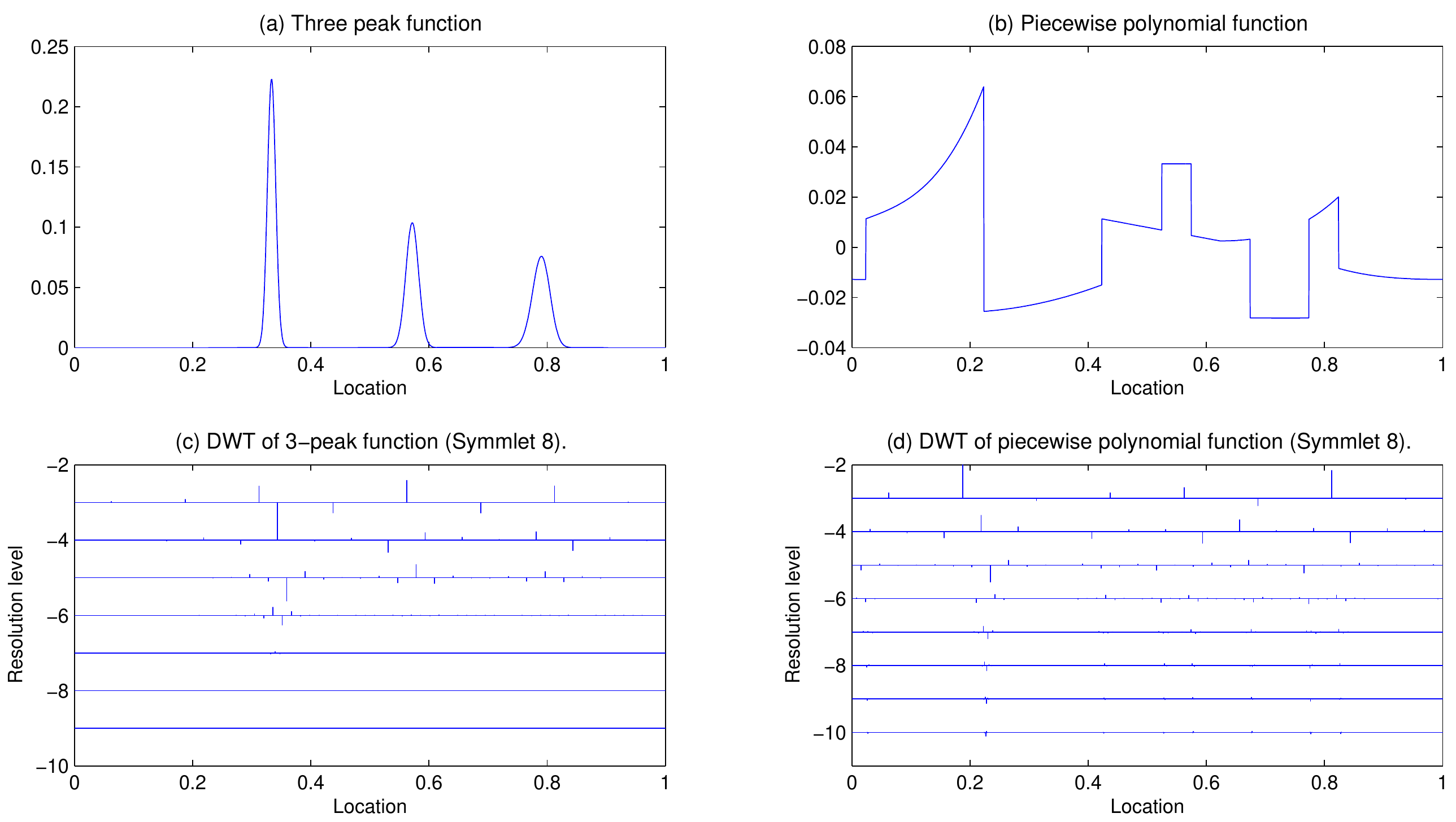}
\caption{(a)~\texttt{peak}: three-peak function evaluated at 1024
  equispaced locations; (b)~\texttt{poly}: piecewise polynomial
  function evaluated at 2048 equispaced locations;
  (c)~\texttt{wc-peak}: discrete wavelet transform (DWT) of the
  three-peak function; (d)~\texttt{wc-poly}: DWT of the piecewise
  polynomial function.  In Plot (c) and (d), each vertical bar is
  proportional in length to the magnitude of the \texttt{Symmlet} 8
  wavelet coefficient at the given location and resolution level.}
\label{fig:rankone}
\end{figure}

For the construction of meaningful singular vectors we use a
functional data analysis context: We choose functions gleaned from the
literature and represent them in wavelet bases where they feature
realistic degrees of sparsity.  In Figure~\ref{fig:rankone}, Plot~(a)
(``\texttt{peak}'') shows the graph of a function with three peaks,
evaluated at $1024$ equispaced locations, while Plot~(b)
(``\texttt{poly}'') shows a piecewise polynomial function, evaluated
at $2048$ equispaced locations.  Both functions create dense
evaluation vectors but sparse wavelet coefficient vectors.  [In all
  simulation results reported below, we use \texttt{Symmelet} 8
  wavelet coefficients \citep{Mallatl09}.]  Multi-resolution plots of
the wavelet coefficients are shown in Plots~(c) (``\texttt{wc-peak}'')
and (d) (``\texttt{wc-poly}'') of Figure~\ref{fig:rankone}.  We choose
$\u_1$ and $\v_1$ to be the wavelet coefficient vectors
\texttt{wc-peak} and \texttt{wc-poly}, respectively.

\begin{table}[tbp] \centering
\begin{small}
\begin{tabular}{c|r|rr|rr|rr|rr}
\hline
losses  & $d_1$ & \multicolumn{2}{|c|}{FIT-SSVD} & \multicolumn{2}{|c|}{LSHM} & \multicolumn{2}{|c|}{PMD-SVD}& \multicolumn{2}{|c}{SVD}\\\hline
&&median&MAD&median&MAD&median&MAD&median&MAD\\\hline
\multirow{3}{*}{$L_{space}(\u_1,\hu_1)$}
& 50&0.0513&0.0009&0.0669&0.0014&0.0783&0.0007&0.5225&0.0034\\
&100&0.0127&0.0003&0.0159&0.0004&0.0254&0.0002&0.1114&0.0005\\
&200&0.0036&0.0001&0.0044&0.0001&0.0102&0.0000&0.0264&0.0001\\\hline
\multirow{3}{*}{$L_{space}(\v_1,\hv_1)$}
& 50&0.0958&0.0008&0.1095&0.0016&0.1399&0.0008&0.6330&0.0025\\
&100&0.0325&0.0004&0.0385&0.0005&0.0566&0.0003&0.1878&0.0006\\
&200&0.0112&0.0001&0.0131&0.0002&0.0241&0.0001&0.0499&0.0001\\\hline
\multirow{3}{*}{$L(\Xi,\hat\Xi)$}
&50&0.1454&0.0014&0.1726&0.0019&0.3280&0.0016&2.2217&0.0082\\
&100&0.0457&0.0004&0.0549&0.0007&0.0973&0.0003&0.3709&0.0009\\
&200&0.0149&0.0001&0.0177&0.0003&0.0364&0.0001&0.0805&0.0002\\\hline
\multirow{3}{*}{$\|\hu_{1}\|_0$}
&50&24&0.1483&22&0.2965&242.5&1.4085&1024&0\\
&100&34&0.1483&32&0.2965&372.5&1.4085&1024&0\\
&200&43&0.1483&41&0.2965&577.0&1.3343&1024&0\\\hline
\multirow{3}{*}{$\|\hv_{1}\|_0$}
&50&18&0.2965&14&0.2965&535.0&2.2239&2048&0\\
&100&40&0.2965&38&0.4448&854.5&1.7050&2048&0\\
&200&66&0.4448&64&0.6672&1303.0&2.0756&2048&0\\\hline
\multirow{3}{*}{time}
&50&0.3364&0.0096&36.7316&0.4497&2.0578&0.0129&1&0\\
&100&0.4401&0.0209&30.8268&0.3305&1.9607&0.0124&1&0\\
&200&0.5685&0.0360&23.7639&0.2542&1.9274&0.0102&1&0\\\hline
\end{tabular}
\caption{Comparison of four methods in the rank-one case: $\u_1$ is
  \texttt{wc-peak}, $\v_1$ is \texttt{wc-poly}, and the noise is iid
  N(0,1).}
\label{table:r1-normal-peak-poly}
\end{small}
\end{table}

For each simulated scenarios, we ran $100$ simulations, applied each
algorithm under comparison, and summarized the results in terms of
median and MAD-based standard error.  The criteria which we use for
comparison of the methods are best explained with reference to
Table~\ref{table:r1-normal-peak-poly}, where we report the results for
iid $N(0,1)$ noise~$Z$:
\begin{itemize}
\item The first block examines the estimation accuracy of the left
  singular vector, with the three rows corresponding to three
  different values of $d_1$. Following \citet{Ma11}, we define the
  loss function for estimating the column space of $U$ for a general
  rank-$r$ by $L_{space}(U,\hU)=\|P_U-P_\hU\|_2^2$ , where $P_U=UU'$
  is the projection matrix onto the subspace spanned by the columns of
  $U$ (which is of size $n \times r$ and has orthonormal columns,
  $U'U=I_r$). In the rank-one case here, the loss reduces to
  $\sin^2\angle(\u_1,\hu_1)$.
\item The second block in Table~\ref{table:r1-normal-peak-poly}
  reports the loss for right singular vectors.
\item The third block shows the scaled recovery error for the low-rank
  signal matrix $\Xi = U D V'$, defined as
  $L(\Xi,\hat\Xi)={\|\hat\Xi-\Xi\|_F^2}/{\|\Xi\|_F^2}$. Here,
  $\hat\Xi=\hU\hD\hV'$ and $\hD=diag(\hd_1,\ldots,\hd_r)$ with
  diagonal entries being $\hd_l=\hu_l'X\hv_l$.
\item The fourth and fifth panels of
  Table~\ref{table:r1-normal-peak-poly} show the sparsity of the
  solutions measured by $\|\hu_1\|_0$ and $\|\hv_1\|_0$, that is, the
  number of nonzero elements in the estimates.
\item The last block shows timing results as a fraction or multiple of
  the ordinary SVD.
\end{itemize}
The results are as follows:
\begin{itemize}
\item From the first three blocks we see that FIT-SSVD uniformly
  outperforms the other methods with respect to the three statistical
  criteria.  While LSHM is not far behind FIT-SSVD, PMD-SVD lags in
  several cases by a factor of two or more.  The ordinary SVD fails
  entirely for low signal strength as the results for $d_1=50$
  illustrate, impressing the need to leverage sparsity in such
  situations.  Rather expectedly, all methods achieve better
  statistical accuracy as the signal strength $d_1$ increases
\item As for the sparsity metrics, FIT-SSVD and LSHM produce similar
  levels of sparsity, while PMD-SVD estimators are much denser.  The
  results also suggest that as the signal strength $d_1$ gets
  stronger, the three sparse SVD methods estimate more coordinates.
\item Finally, the timing results indicate that FIT-SSVD is faster
  than all other methods, the ordinary SVD included.  LSHM stands out
  as slower than FIT-SSVD by factors of over 40 to over 100.  PMD-SVD
  is more competitive but still at least a factor of three slower than
  FIT-SSVD.  The variation in time for PMD-SVD is small because the
  majority is spent in cross-validation.
\end{itemize}

To examine the effect of heavy-tailed noise, we report in
Table~\ref{table:r1-t5-peak-poly} the simulation results when the
entries of the noise matrix $Z$ are distributed iid $\sqrt{3/5}\,t_5$,
all else being the same as in Table~\ref{table:r1-normal-peak-poly}.
[Recall that the scaling factor $\sqrt{3/5}$ is used to ensure unit
  variance.]  The statistical performance for all methods is worse
than in Table~\ref{table:r1-normal-peak-poly}.  In terms of the
statistical metrics, the performances of FIT-SSVD and LSHM are in a
statistical dead heat, whereas PMD-SVD trails behind by as much as a
factor of two in the case of high signal strength, $d_1=200$.  Again,
FIT-SSVD and LSHM have comparable sparsities, whereas PMD-SVD is much
denser.  In terms of computation time, again FIT-SSVD is uniformly
fastest, followed by PMD-SVD which trails by factors of over two to
over five, and LSHM being orders of magnitude slower (by factors of 28
to~110).

\begin{table}[tbp] \centering
\begin{small}
\begin{tabular}{c|r|rr|rr|rr|rr}
\hline
losses  & $d_1$ & \multicolumn{2}{|c|}{FIT-SSVD} & \multicolumn{2}{|c|}{LSHM} & \multicolumn{2}{|c|}{PMD-SVD}& \multicolumn{2}{|c}{SVD}\\\hline
&&median&MAD&median&MAD&median&MAD&median&MAD\\\hline
\multirow{3}{*}{$L_{space}(\u_1,\hu_1)$}
& 50&0.0802&0.0015&0.0819&0.0017&0.0907&0.0011&0.5405&0.0037\\
&100&0.0177&0.0003&0.0180&0.0004&0.0282&0.0003&0.1115&0.0006\\
&200&0.0048&0.0001&0.0047&0.0001&0.0108&0.0001&0.0262&0.0002\\\hline
\multirow{3}{*}{$L_{space}(\v_1,\hv_1)$}
& 50&0.1193&0.0014&0.1191&0.0018&0.1560&0.0014&0.6432&0.0039\\
&100&0.0451&0.0005&0.0415&0.0007&0.0601&0.0003&0.1870&0.0006\\
&200&0.0145&0.0002&0.0137&0.0002&0.0249&0.0001&0.0498&0.0002\\\hline
\multirow{3}{*}{$L(\Xi,\hat\Xi)$}
& 50&0.1944&0.0024&0.1937&0.0028&0.3719&0.0028&2.2624&0.0107\\
&100&0.0625&0.0007&0.0600&0.0009&0.1041&0.0005&0.3706&0.0011\\
&200&0.0192&0.0002&0.0187&0.0002&0.0378&0.0001&0.0805&0.0002\\\hline
\multirow{3}{*}{$\|\hu_{1}\|_0$}
& 50&20&0.2965&21.5&0.3706&235.0&1.5567&1024&0\\
&100&31&0.1483&33.0&0.2965&364.0&1.8532&1024&0\\
&200&40&0.1483&41.0&0.2965&569.5&2.0015&1024&0\\\hline
\multirow{3}{*}{$\|\hv_{1}\|_0$}
& 50&13&0.1483&14.0&0.2965&526.0&2.0015&2048&0\\
&100&31&0.2965&38.5&0.5189&841.5&2.1498&2048&0\\
&200&56&0.2965&64.5&0.6672&1307.5&2.2980&2048&0\\\hline
\multirow{3}{*}{time}
& 50&0.3714&0.0190&41.0695&0.9339&2.0826&0.0046&1&0\\
&100&0.8072&0.0652&30.5216&0.3774&2.0187&0.0039&1&0\\
&200&0.8238&0.0710&23.0527&0.2554&1.9520&0.0048&1&0\\\hline
\end{tabular}
\caption{Comparison of four methods in the rank-one case: $\u_1$ is
  \texttt{wc-peak}, $\v_1$ is \texttt{wc-poly}, and the noise is
  iid~${\scriptstyle \sqrt{3/5}}\,t_5$.}
\label{table:r1-t5-peak-poly}
\end{small}
\end{table}


\subsection{Rank-two results}
\label{ssec:simulation-r2}

\begin{figure}[htp]
\centering
\includegraphics[width=0.9\textwidth]{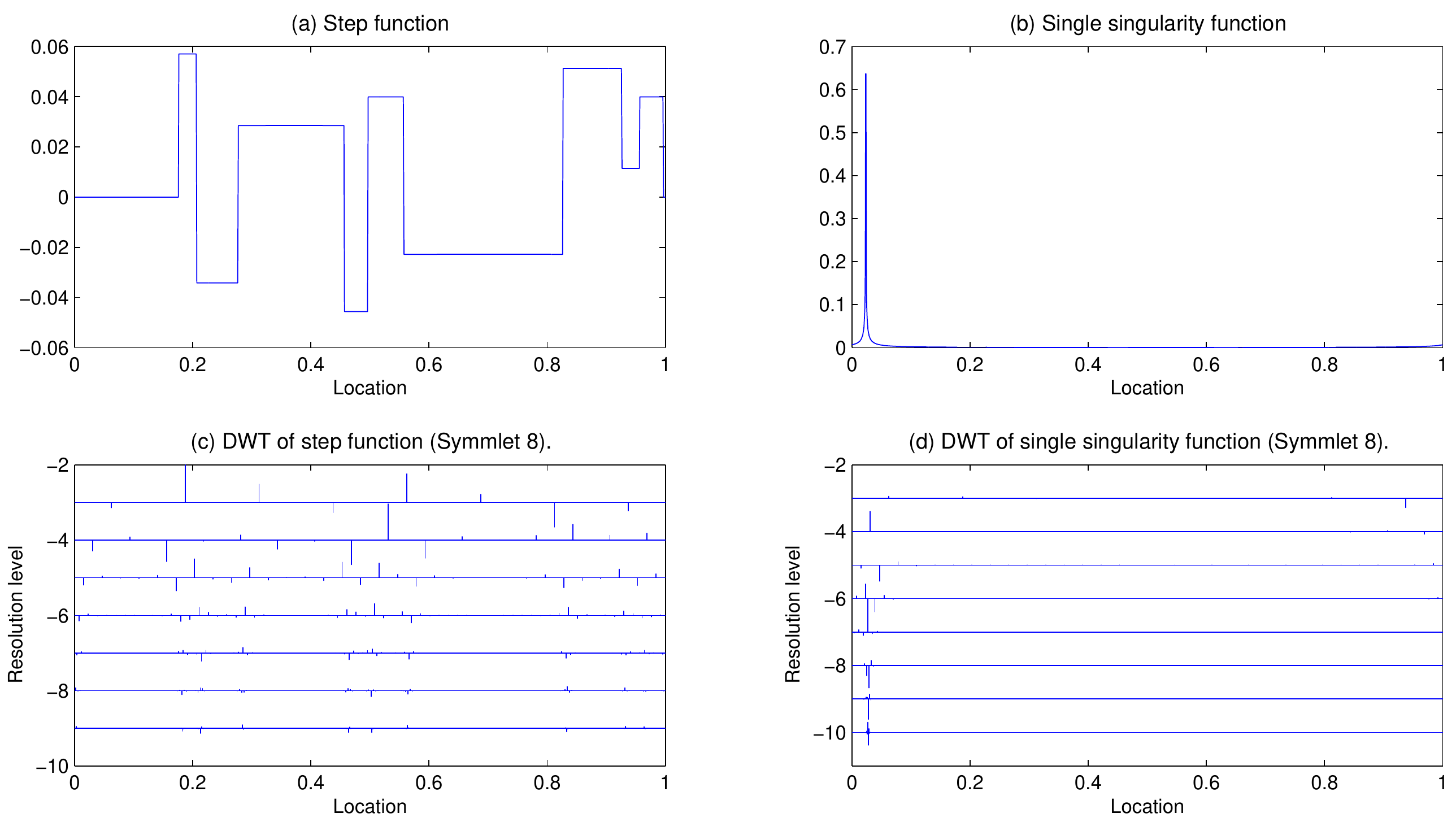}
\caption{(a)\texttt{step}: step function evaluated at 1024 equispaced
  locations, (b)\texttt{sing}: single singularity function evaluated
  at 2048 equispaced locations, (c)\texttt{wc-step}: DWT of step
  function, (d)\texttt{wc-sing}: DWT of single singularity function.}
\label{fig:ranktwo}
\end{figure}

We show next simulation results for data according to model
\eqref{eq:model} with $r = 2$, and again $n = 1024$ and $p = 2048$.
The singular values $(d_1, d_2)$ range among the pairs $(100,50)$,
$(200,50)$, and $(200,100)$.  The singular vectors are $\u_1 = $
\texttt{wc-peak}, $\v_1 = $\texttt{wc-poly}, $\u_2 = $
\texttt{wc-step}, and $\v_2 = $ \texttt{wc-sing}, the properties of
the latter two vectors being shown in Figure~\ref{fig:ranktwo}.

Table~\ref{table:r2} reports the results from $100$ repetitions when
the noise is iid~$N(0,1)$.  In terms of statistical metrics, FIT-SSVD
always outperforms LSHM though not hugely.  PMD-SVD does slightly
better than FIT-SSVD for $L_{space}(U,\hU)$, but much worse for
$L_{space}(V,\hV)$ and $L(\Xi,\hat\Xi)$.  This is due to the special
type of cross-validation used in the package \texttt{PMA}: the
parameters $s_u, s_v$ are set to be proportional to each other after
being scaled according to the dimensionality, $\sqrt{n}$ and
$\sqrt{p}$, which essentially reduces the simultaneous
cross-validation on two parameters to one.  Therefore, PMD-SVD
actually enforces the same level of sparsity on $\hu$ and $\hv$.

In terms of sparsity of the estimators, the fourth and fifth blocks
show the cardinality of the joint support of the estimated singular
vectors, which indicate that FIT-SSVD and LSHM are again about
comparable, and PMD-SVD is much denser as in the rank-one case.  [We
  do not compare the losses and the $l_0$ norms for individual
  singular vectors because LSHM and PMD-SVD do not produce orthogonal
  singular vectors.]

Finally, in terms of computation time, FIT-SSVD dominates again, and
the differences become somewhat more pronounced than in the rank-one
case.  In the high signal scenario, $d_1=200$ and $d_2=100$, FIT-SSVD
gets a boost because by avoiding the costly bootstrap in
Algorithm~\ref{algo:threshold-level} because
Condition~\ref{algo:step:error-if-condition} is satisfied and the much
cheaper normal approximation on Line~\ref{algo:step:error-normal} of
Algorithm~\ref{algo:threshold-level} can be used to compute the
threshold level.  Since LSHM repeats its scheme on the residual matrix
to get the second layer of SVD, computation time doubles.  As for
PMD-SVD, since the time is mainly spent in cross-validation and the
same penalty parameter is used for different ranks, the increase in
time is not obvious.

\begin{table}[tbp] \centering
\begin{small}
\begin{tabular}{c|rr|rr|rr|rr|rr}
\hline
losses  & $d_1$ & $d_2$ & \multicolumn{2}{|c|}{FIT-SSVD} & \multicolumn{2}{|c|}{LSHM} & \multicolumn{2}{|c|}{PMD-SVD}& \multicolumn{2}{|c}{SVD}\\\hline
&&&median&MAD&median&MAD&median&MAD&median&MAD\\\hline
\multirow{3}{*}{$L_{space}(U,\hU)$}
&100& 50&0.1163&0.0010&0.1413&0.0021&0.1022&0.0009&0.5315&0.0037\\
&200& 50&0.1148&0.0013&0.1422&0.0018&0.1007&0.0009&0.5265&0.0027\\
&200&100&0.0376&0.0003&0.0443&0.0006&0.0321&0.0003&0.1114&0.0005\\\hline
\multirow{3}{*}{$L_{space}(V,\hV)$}
&100& 50&0.0514&0.0009&0.0596&0.0010&0.1230&0.0008&0.6376&0.0029\\
&200& 50&0.0506&0.0009&0.0601&0.0011&0.1259&0.0006&0.6293&0.0023\\
&200&100&0.0144&0.0002&0.0172&0.0003&0.0538&0.0002&0.1870&0.0005\\\hline
\multirow{3}{*}{$L(\Xi,\hat\Xi)$}
&100& 50&0.0691&0.0006&0.0825&0.0007&0.1403&0.0004&0.7439&0.0017\\
&200& 50&0.0234&0.0001&0.0285&0.0002&0.0529&0.0001&0.2070&0.0005\\
&200&100&0.0228&0.0001&0.0261&0.0002&0.0483&0.0001&0.1387&0.0003\\\hline
\multirow{3}{*}{$|supp(\hu_1)\cup supp(\hu_2)|$}
&100& 50&49&0.2965&45.0&0.4448&479&1.1861&1024&0\\
&200& 50&56&0.2965&49.0&0.2965&649&1.5567&1024&0\\
&200&100&77&0.2965&73.5&0.5189&657&1.3343&1024&0\\\hline
\multirow{3}{*}{$|supp(\hv_1)\cup supp(\hv_2)|$}
&100& 50&54&0.2965&50.5&0.3706&1158.0&2.2239&2048&0\\
&200& 50&78&0.2965&74.5&0.5189&1486.5&2.1498&2048&0\\
&200&100&81&0.4448&82.5&0.5930&1623.0&2.1498&2048&0\\\hline
\multirow{3}{*}{time}
&100& 50&1.1675&0.0829&64.7840&0.6037&2.7991&0.0141&1&0\\
&200& 50&1.4572&0.1011&55.6839&0.5436&2.7018&0.0142&1&0\\
&200&100&0.8000&0.0668&54.2361&0.2363&2.6429&0.0073&1&0\\\hline
\end{tabular}
\caption{Comparison of four methods for the rank-two case, with noise
  iid~$N(0,1)$.}
\label{table:r2}
\end{small}
\end{table}



\section{Real data examples}
\label{sec:data}

All the methods mentioned above require sparse singular vectors (with
most entries close to zero).  One source of such data is two-way
functional data whose row and column domains are both structured, for
example, temporally or spatially, as when the data are time series
collected at different locations in space.  Two-way functional data
are usually smooth as functions of the row and column domains.  Thus,
if we expand them in suitable basis functions, such as an orthonormal
trigonometric basis, the coefficients should be sparse
\citep{Johnstone11}.


\subsection{Mortality rate data}
\label{ssec:data-mortality}

As our first example we use the US mortality rate data from the
Berkeley Human Mortality Database (http://www.mortality.org/).  They
contain mortality rates in the United States for ages 0 to 110 from
1933 to 2007.  The data for people older than 95 was discarded because
of their noisy nature.  The matrix $X$ is of size $96 \times 75$, each
row corresponding to an age group and each column to a one-year
period.  We first pre- and post- multiply the data matrix with
orthogonal matrices whose columns are the eigenvectors of second order
difference matrices of proper sizes; the result is a matrix of
coefficients of the same size as $X$.  The rank of the signal is
estimated to be 2 using bi-crossvalidation (Section~\ref{ssec:rank}).
We then applied FIT-SSVD, LSHM, PMD-SVD and ordinary SVD to get the
first two pairs of singular vectors.  Finally, we transformed the
sparse estimators of the singular vectors back to the original basis
to get smooth singular vectors.

\begin{figure}[htp]
\centering
\includegraphics[width=0.95\textwidth]{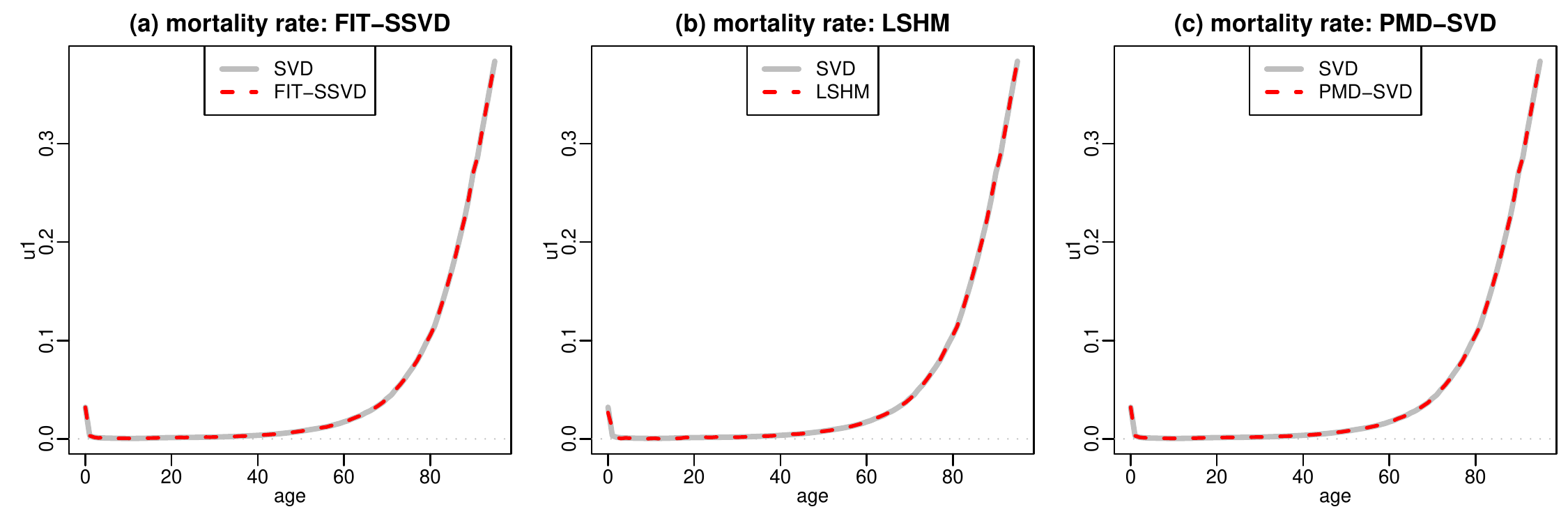}
\caption{Mortality data: plot of $\hu_1$.
Panel (a): FIT-SSVD vs. SVD;
Panel (b): LSHM vs. SVD;
Panel (c): PMD-SVD vs. SVD.}
\label{fig:age1}
\end{figure}

\begin{figure}[htp]
\centering
\includegraphics[width=0.95\textwidth]{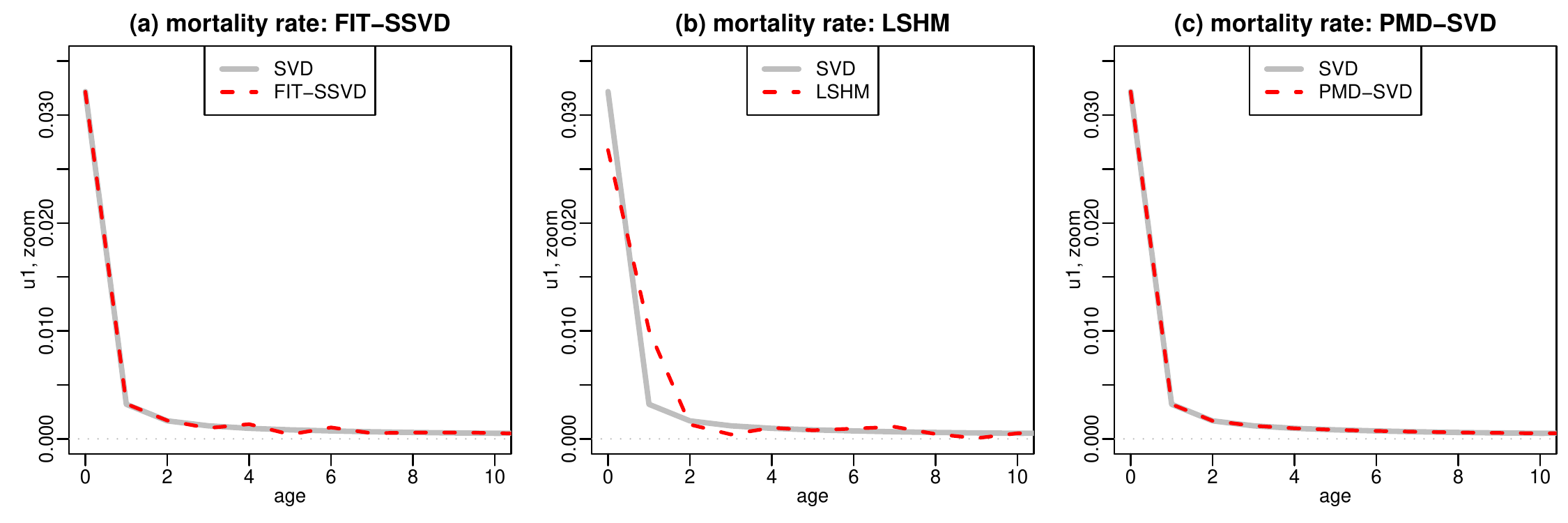}
\caption{Mortality data: Plot of $\hu_1$.  Zoom of the lower left
  corner of Figure \ref{fig:age1}.  Everything else is the same as in
  Figure~\ref{fig:age1}.}
\label{fig:age1-zoom}
\end{figure}

\begin{figure}[htp]
\centering
\includegraphics[width=0.95\textwidth]{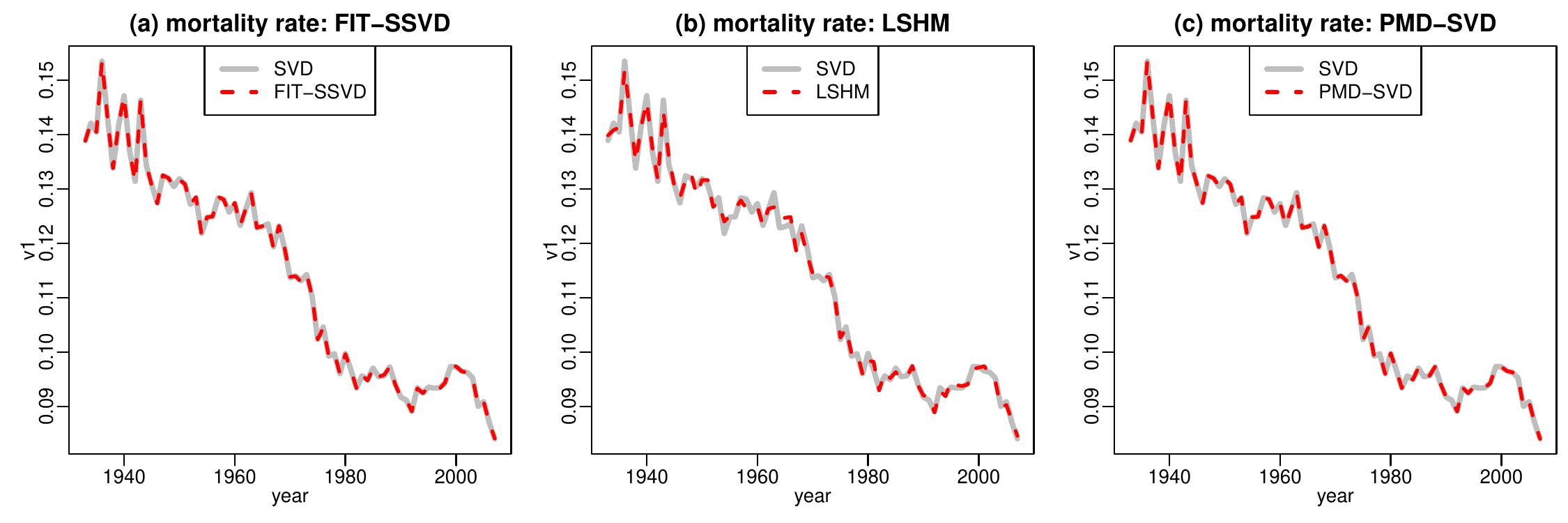}
\caption{Mortality data: Plot of $\hv_1$.  Everything else is the same
  as in Figure~\ref{fig:age1}.}
\label{fig:year1}
\end{figure}

\begin{figure}[htp]
\centering
\includegraphics[width=0.95\textwidth]{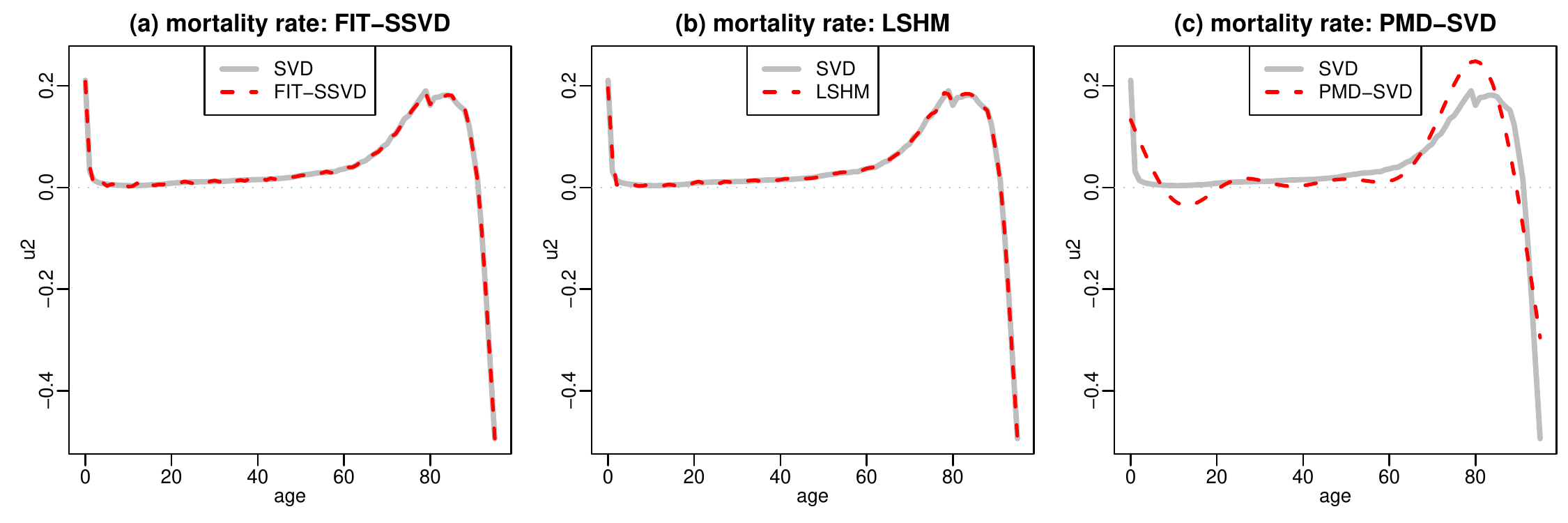}
\caption{Mortality data: plot of $\hu_2$. Everything else is the same as Figure \ref{fig:age1}.}
\label{fig:age2}
\end{figure}

\begin{figure}[htp]
\centering
\includegraphics[width=0.95\textwidth]{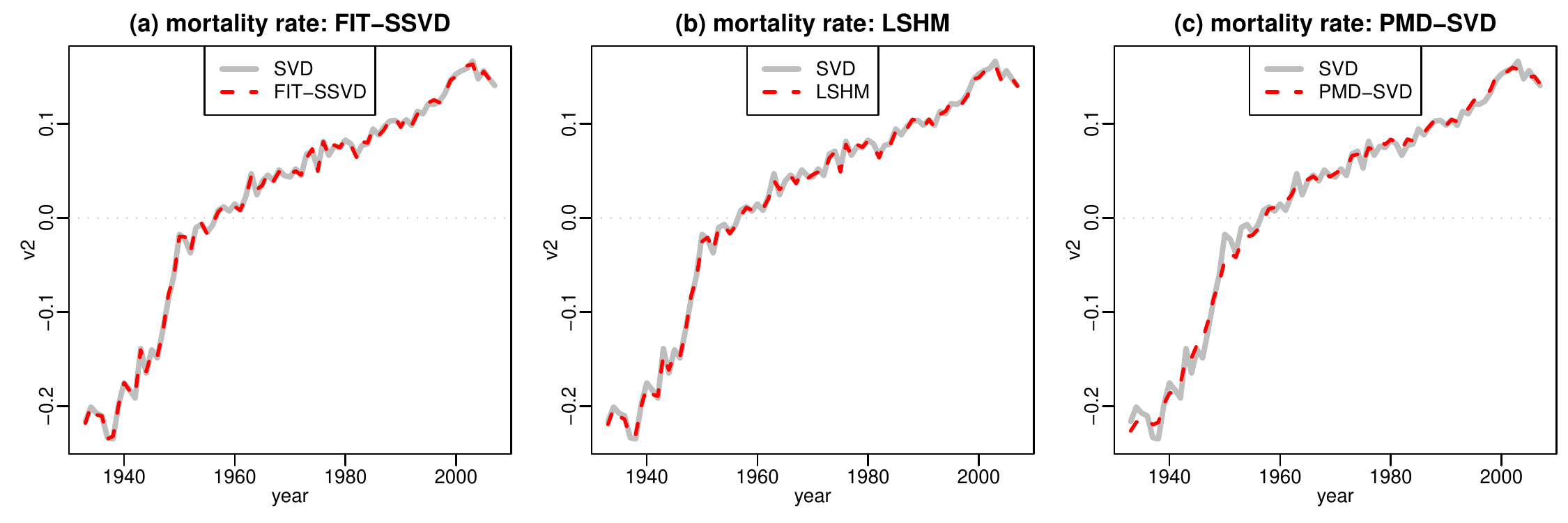}
\caption{Mortality data: plot of $\hv_2$. Everything else is the same as Figure \ref{fig:age1}.}
\label{fig:year2}
\end{figure}

The estimated number of nonzero elements in each singular vector
(before the back transformation) is summarized in
Table~\ref{table:mortality}: none gives very sparse solutions.  This
is reasonable, because the mortality rate data is of low noise and for
data with no noise we should just use the ordinary SVD.  Because this
data is of small size, it only takes a few seconds for all the
algorithms.  The plot of singular vectors for all the methods are
shown in Figures~\ref{fig:age1} to \ref{fig:year2}.  The red dashed
line in the left plot is for FIT-SSVD, in the middle for LSHM, and on
the right for PMD-SVD.  We use the wider gray curve for the ordinary
SVD as a reference.

\begin{table}[tbp] \centering
\begin{tabular}{c|cccc}
\hline
& FIT-SSVD & LSHM & PMD-SVD & SVD\\\hline
$\|\hu_{1}\|_0$&82&48&96&96\\
$\|\hu_{2}\|_0$&86&56&7&96\\\hline
$\|\hv_{1}\|_0$&66&45&75&75\\
$\|\hv_{2}\|_0$&70&45&43&75\\\hline
\end{tabular}
\caption{Mortality data: number of nonzero coordinates in the transformed domain for four methods.}
\label{table:mortality}
\end{table}

Figure~\ref{fig:age1} shows the first left singular vector plotted
against age.  The curve $\hu_1$ shows a pattern for mortality as a
function of age: a sharp drop between age 0 and 2, then a gradual
decrease till the teen years, flat till the 30s, after which begins an
exponential increase.  Figure~\ref{fig:age1-zoom} zooms into the lower
left corner of Figure~\ref{fig:age1} to show the details between age 0
and 10.  LSHM, as always turns out to be the sparsest (or smoothest)
among the three iterative procedures in the transformed (or original)
domain.  We believe that FIT-SSVD and PMD-SVD make more sense based on
a parallel coordinates plot of the raw data (not shown here), in which
the drop in the early age appears to be sharp and therefore should not
be smoothed out.  Figure~\ref{fig:year1} shows the first right
singular vectors plotted against year.  It implies that mortality
decreases with time.  All of the panels show a wiggly structure, with
LSHM again being the smoothest.  Here, too, we believe that the zigzag
structure is real and not due to noise in the raw data, based again on
a parallel coordinate plot of the raw data.  The zigzags may well be
systematic artifacts, but they are unlikely to be noise.

The second pair of singular vectors is shown in Figures~\ref{fig:age2}
and \ref{fig:year2}: They correct the pattern that the first pair of
singular vectors does not capture.  The contrast mainly focuses on
people younger than 2 or between 60 and 90 where $\hu_2$ is positive.
Also, $\hv_2$ has extreme negative or positive values towards the both
ends, 1940s and 2000s.  Together, they suggest that babies and older
people had lower mortality rates in the 1940s and higher mortality
rates in the 2000s than what the first component expresses.  One final
aspect to note is the strange behavior of $\hu_{2,PMD-SVD}$, recalling
that $\hu_{1,PMD-SVD},\hv_{1,PMD-SVD},\hv_{2,PMD-SVD}$ all follow the
ordinary SVD very closely.  We think this is again due to the
cross-validation technique they use.

\citet{Huang09} also used the mortality data from 1959 to 1999 to
illustrate their version of regularized SVDs to get smooth singular
vectors by adding second order difference penalties.  If we compare
the results shown in this section with theirs, our solutions lack the
smoothness of their solutions, but we think we recover more
information from the data by capturing not only the general trend but
also local details such as year-to-year fluctuations.


\subsection{Cancer data}
\label{ssec:data-cancer}

We consider next another data example where some sparse structure may
be expected to exist naturally.  The cancer data used by
\citet{Lee+10} (who in turn have them from \citet{Liu+08}) consists of
the gene expression levels of 12,625 genes for 56 cases of four
different types of cancer.  It is believed that only a part of the
genes regulate the types and hence the singular vectors corresponding
to the genes should ideally be sparse.  We apply the four SVD methods
directly to the raw data without change of basis.

Before we proceed it may be proper to discuss briefly some modeling
ambiguities posed by this dataset as it is not a priori clear whether
a PCA or SVD model is more appropriate.  It might be argued that the
cases really should be considered as being sampled from a population,
hence PCA would be the proper analysis, with the genes representing
the variables.  The counter argument is, however, that the cases are
stratified, and the strata are pure convenience samples of sizes that
bear no relation to the sizes of naturally occurring cancer
populations.  A dual interpretation with genes as samples and cases as
variables would be conceivable also, but it seems even more far
fetched in the absence of any sampling aspect with regard to genes.
In light of the problems raised by any sampling assumption, it would
seem more appropriate to condition on the cases and the genes and
adopt a fixed effects view of the data.  As a result the SVD model
seems less problematic than either of the dual PCA models.

We first attempted to estimate the rank of the signal using
bi-crossvalidation (Section~\ref{ssec:rank}), but it turns out that
the rank is sensitive to the choice of $\alpha$ (Holm family-wise
error) and $\beta$ (Huberization quantile) in
Algorithm~\ref{algo:init}, ranging from $r$=3 to $r$=5.  We decided to
use $r=3$ because this is the number of contrasts required to cluster
the cases into four groups.  Also, this is the rank used by
\citet{Lee+10}, which grants comparison of their and our results.

On a different note, running LSHM on these data with rank three took a
couple of hours, which may be a disincentive for users to seek even
higher ranks.  The hours of run time of LSHM compares with a few
minutes for PMD-SVD and merely a few seconds for FIT-SSVD.  (In
addition, LSHM's third singular vectors do not seem to converge within
300 iterations.)

Table~\ref{table:cancer} summarizes the cardinalities of the union of
supports of three singular vectors for each method.  For the
estimation of left singular vectors corresponding to different genes,
the PMD-SVD solution is undesirably dense, while FIT-SSVD and LSHM
give similar levels of sparsity.  For the estimation of right singular
vectors corresponding to the cases, we would expect that all cases
have their own effects rather than zero, so it is not surprising that
the estimated singular vectors are dense.

\begin{table}[tbp] \centering
\begin{tabular}{c|cccc}
\hline
& FIT-SSVD & LSHM & PMD-SVD & SVD\\\hline
$|\cup_{l=1}^3\supp(\hu_l)|$&4688&4545&12625&12625\\
$|\cup_{l=1}^3\supp(\hv_l)|$&56&56&54&56\\\hline
\end{tabular}
\caption{Cancer data: summary of cardinality of joint support of three
  singular vectors for four methods.}
\label{table:cancer}
\end{table}

\begin{figure}[htp]
\centering
\includegraphics[width=0.8\textwidth]{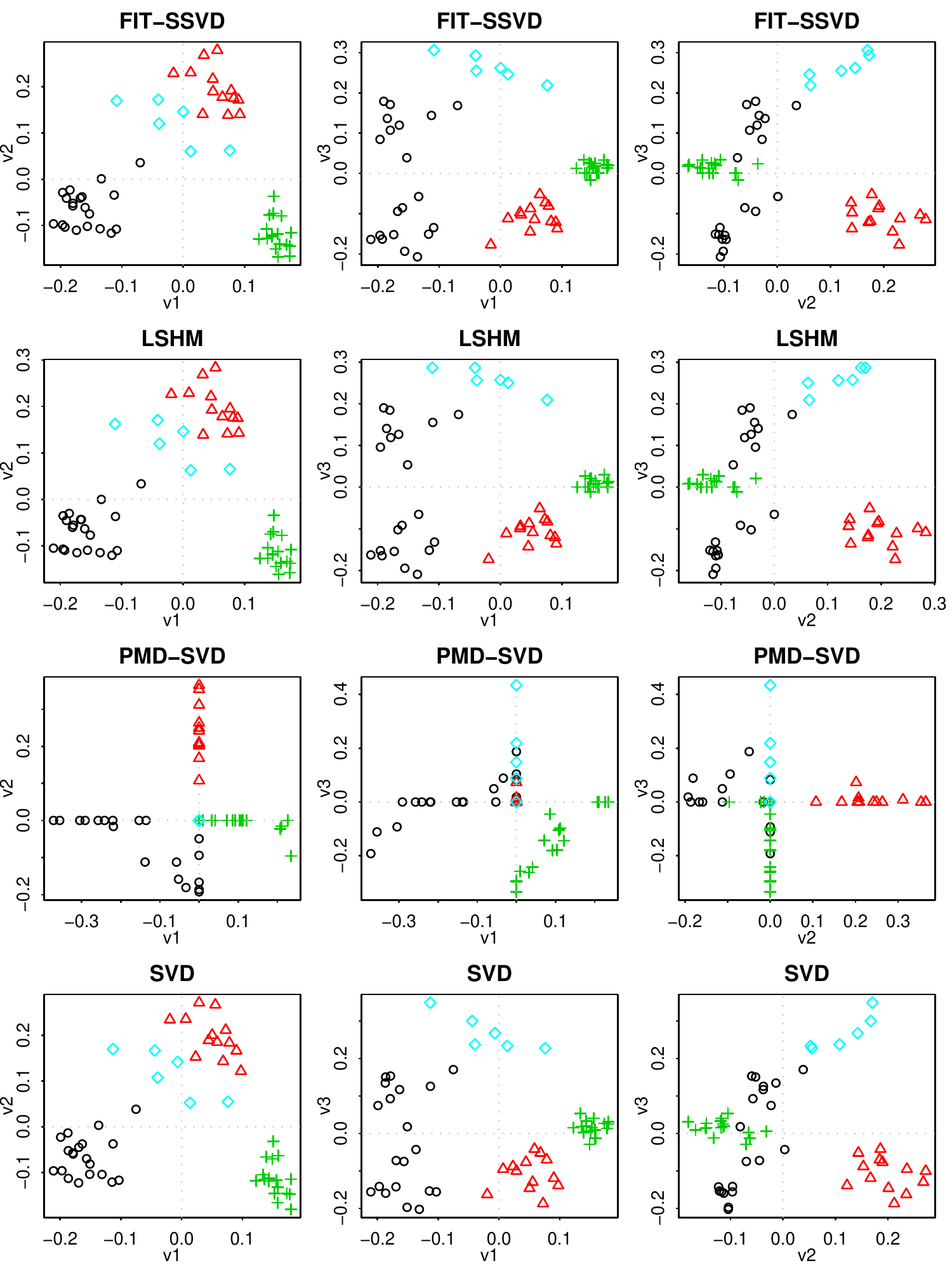}
\caption{Cancer data: Scatterplots of the entries of the first three
  right singular vectors $\hv_l, l=1,2,3$ for four methods.  Points
  represent patients.  Black circle: Carcinoid; Red triangle: Colon;
  Green cross: Normal; Blue diamond: SmallCell.}
\label{fig:cancer-scatterplot}
\end{figure}

Figure~\ref{fig:cancer-scatterplot} shows the scatterplots of the
entries of the first three right singular vectors for the four
methods.  Points represent patients, each row represents one method,
and each column corresponds to two of the three singular vectors.  The
four known groups of patients are easily discerned in the plots.  A
curiosity is the cross-wise structure produced by PMD-SVD, where the
singular vectors are nearly mutually exclusive: if one coordinate in a
singular vector is non-zero, most corresponding coordinates in the
other singular vectors are zero.  The other three methods, including
the ordinary SVD, agree strongly among each other in the placement of
the patients.  The agreement with the ordinary SVD is not a surprise
as $p=56$ is a relatively small column dimension on which sparsity may
play a less critical role compared to the row dimension with
$n=12625$.  Yet, the three sparse methods give clearer evidence that
the carcinoid group (black cirlces) falls into two subgroups than the
ordinary SVD.  According to FIT-SSVD and LSHM the separation is along
$\hv_3$ (center and right hand plots), whereas according to PMD-SVD it
is by lineup with $\hv1$ and $\hv2$, respectively (left hand plot).

\begin{figure}[htp]
\centering
\includegraphics[width=0.8\textwidth]{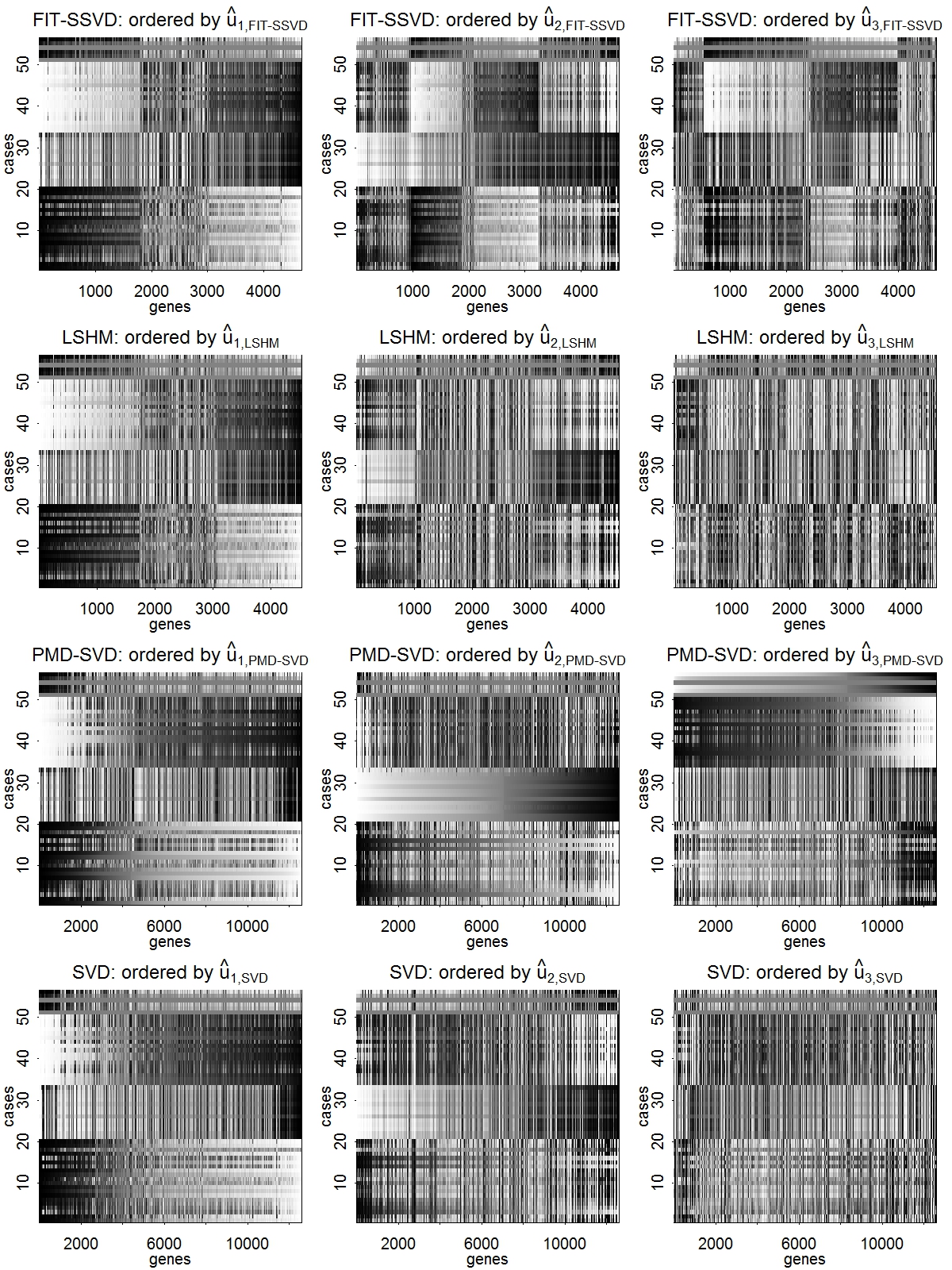}
\caption{Cancer data: Image plots of the rank-three approximations
  $\sum_{l=1,2,3} \hd_l \hu_l \hv_l'$ whose values are gray-coded.
  Each image is laid out as cases (=~rows) by genes (=~columns).  The
  same rank-three approximation is shown three times for each method
  (left to right), each time sorted according to a different $\hu_l$
  ($l=1,2,3$).  (The mapping of the rank-three approximation values to
  gray scales is by way of a rank transformation, using a separate
  transformation for each image.  Rank transformations create
  essentially uniform distributions that better cover the range of
  gray scale values.)  }
\label{fig:cancer-checkboard}
\end{figure}

Figure~\ref{fig:cancer-checkboard} shows checkerboard plots of the
reconstructed rank-three approximations, layed out with patients on
the vertical axis and genes on the horizontal axis.  Each row of plots
represents one method, and the plots in a given row show the same
reconstructed matrix but successively ordered according to the
coordinates of the estimated left singular vectors $\hu_1$, $\hu_2$
and $\hu_3$.  There are fewer than 5000 genes shown for FIT-SSVD and
LSHM, because the rest are estimated to be zero, whereas all 12,625
genes are shown for PMD-SVD and SVD (Table~\ref{table:cancer}).  We
can see clear checkerboard structure in some of the plots, indicating
biclustering.  In spite of the strong similarity between the patient
projections for FIT-SSVD and LSHM in
Figure~\ref{fig:cancer-scatterplot}, there is a clear difference
between these methods in the reconstructions in
Figure~\ref{fig:cancer-checkboard}: The FIT-SSVD solution exhibits the
strongest block structure in its $\hu_2$-based sort (center plot, top
row), implying the strongest evidence of clustering among its
non-thresholded genes.  Since these blocks consist of many hundreds of
genes, it would surprisingly suggest that the differences between the
four patient groups run into the hundreds, not dozens, of genes.

In spite of the differences in checkerboard patterns in
Figure~\ref{fig:cancer-checkboard}, the three left singular vectors
are highly correlated between FIT-SSVD and LSHM: $\corr$ = 0.985,
0.981, and 0.968, respectively, and the top 20 genes with largest
magnitude in the estimated three left singular vectors of FIT-SSVD
overlap with those of LSHM except for one gene in the second singular
vector.
These shared performance aspects notwithstanding, the two methods
differ hugely in computing time, FIT-SSVD taking seconds, LSHM taking
a couple of hours.



\section{Discussion}
\label{sec:discussion}

We presented a procedure, called FIT-SSVD, for the fast and
simultaneous extraction of singular vectors that are sparse in both
the row and the column dimension.  While the procedure is state of the
art in terms of statistical performance, its overriding advantage is
sheer speed.  The reasons why speed matters are several: (1)~Faster
algorithms enable the processing of larger datasets.  (2)~The use of
SVDs in data analysis is most often for exploratory ends which call
for unlimited iterations of quickly improvised steps --- something
that is harder to achieve as datasets grow larger.  (3)~Sparse
multivariate technology is still a novelty and hence at an
experimental stage; if its implementation is fast, early adopters of
the technology have a better chance to rapidly gain experience by
experimenting with its parameters.  (4)~If a statistical method such
as sparse SVD has a fast implementation, it can be incorporated as a
building block in larger methodologies, for example, in processing
data arrays that are more than two-dimensional.  For these reasons we
believe that fast SVD technology is of the essence for its success.

A unique opportunity for sparse approaches is to achieve faster speed
than standard non-sparse approaches when the structure in the data is
truly sparse.  Our algorithm achieves this to some extent through
initialization that is both sparse and smart: sparse initialization
consists of a standard SVD of smaller size than the full data matrix,
while smart (in particular: non-random) initialization reduces the
number of iterations to convergence.  A statistical benefit is that
inconsistent estimation by the standard SVD on large data matrices
with weak signal is avoided.  ---~An imperative for fast
implementations is avoiding where possible such slow devices as
cross-validation.  A considerable speed advantage we achieve is
through relatively fast (non-crossvalidated) selection of thresholding
levels based on an analytical understanding of their function.

Our proposal has conceptual and theoretical features that are unique
at this stage of the development of the field: (1)~FIT-SSVD extracts
$r$ orthogonal left- and right-singular vectors simultaneously, which
puts it more in line with standard linear dimension reduction where
orthogonal data projections are the norm.  In addition, simultaneous
extraction can be cast as subspace extraction, which provides a degree
of immunity to non-identifiability and slow convergence of individual
singular vectors when some of the first $r$ underlying singular values
are nearly tied: since we measure convergence in terms of distance
between successive $r$-dimensional subspaces, our algorithm does not
need to waste effort in pinning down ill-determined singular vectors
as long as the left- and right-singular subspaces are well-determined.
Such a holistic view of the rank-$r$ approximation is only available
to simultaneous but not to successive extraction.  (2)~FIT-SSVD is
derived from asymptotic theory that preceded its realization as a
methodology: For Gaussian noise in the model (\ref{eq:model}), we
\citep{Yang+11} showed that our algorithm with appropriately chosen
parameters achieves the rate of the minimax lower bound.  In other
words, in a specific parameter space, our algorithm is asymptotically
optimal in terms of minimizing the maximum possible risk over the
whole parameter space.

As for future work, the current state of the art raises several
questions.  For one, it would be of interest to better understand the
relative merits of the currently proposed sparse SVD approaches since
they have essential features in common, such as power iterations and
thresholding.  Another natural question arises from the fact that
sparse SVDs build on the sequence model: many methods for choosing
parameters from the data have been shown to be asymptotically
equivalent to first order in the sequence model (see, e.g., Haerdle et
al.~(1988)), including cross-validation, generalized cross-validation,
Rice's method based on unbiased estimates of risk, final prediction
error, and the Akaike information criterion.  Do these asymptotic
equivalences hold in the matrix setting for sparse SVD approaches?
How does the choice of the BIC in LSHM compare?  Also, our algorithm
and underlying theory allow a wide range of thresholding functions: Is
there an optimal choice in some sense?  Further, there exists still a
partial disconnect between asymptotic theory and practical
methodology: The theory requires a strict rank~$r$ model, whereas by
all empirical evidence the algorithm works well in a ``trailing rank''
situation where real but small singular values exist.  Finally, there
is a robustness aspect that is specific to sparse SVD approaches:
heavier than normal tails in the noise distribution generate ``random
factors'' caused by single outlying cells.  While we think we have
made reasonable and empirically successful choices in drawing from the
toolkit of robustness, we have not provided a theoretical framework to
justify them.  ---~Just the same, even if the proposed FIT-SSVD
algorithm may be subject to some future tweaking, in the substance it
has the promise of lasting merit.

\bibliographystyle{plainnat}

\bibliography{svdsparse}


\end{document}